\documentstyle[12pt,psfig,epsf,epsfig]{article}
\newcommand {\be}{\begin{equation}}
\newcommand {\ee}{\end{equation}}
\newcommand {\cE}{{\cal E}}

\newcommand {\cn}{{\cal N}}

\begin{document}

\begin{center}
{\Large\bf Critical Unmixing of Polymer Solutions} \\[1 cm]

{\large\bf Helge Frauenkron$^1$ and Peter Grassberger$^{1,2}$}\\[0.5cm]
 
$^1$ HLRZ c/o Forschungszentrum J\"ulich, D-52425 J\"ulich, Germany\\
$^2$ Physics Department, University of Wuppertal, D-42097 Wuppertal, Germany
\\[0.5cm]

\bf{\today}
\end{center}

\begin{abstract}

We present Monte Carlo simulations of semidilute solutions 
of long self-attracting chain polymers near their Ising type critical 
point. The polymers are modeled as monodisperse 
self-avoiding walks on the simple cubic 
lattice with attraction between non-bonded nearest neighbors. Chain 
lengths are up to $N=2048$, system sizes are up to $2^{21}$ lattice sites 
and $2.8\times 10^5$ monomers. These simulations used the recently 
introduced pruned-enriched Rosenbluth method which proved extremely 
efficient, together with a histogram method for estimating finite 
size corrections. Our most clear result is that chains at the critical 
point are Gaussian for large $N$, having end-to-end distances 
$R\sim\sqrt{N}$. Also the distance $T_\Theta-T_c(N)$ (where $T_\Theta =
\lim_{N\to\infty} T_c(N)$) scales with the mean field exponent, $T_\Theta
-T_c(N)\sim 1/\sqrt{N}$. The critical density {\it seems} to scale 
with a non-trivial exponent similar to that observed in experiments. 
But we argue that this is due to large logarithmic corrections.
These corrections are similar 
to the very large corrections to scaling seen in recent analyses of 
$\Theta$-polymers, and qualitatively predicted by the field theoretic 
renormalization group. The only serious
deviation from this simple global picture concerns the 
$N$-dependence of the order parameter amplitudes which disagrees 
with a minimalistic ansatz of de Gennes. But this might be due to 
problems with finite size scaling. We find that the
finite size dependence of the density of states $P(E,n)$ (where $E$ 
is the total energy and $n$ is the number of chains) is slightly but 
significantly different from that proposed recently by several authors.

\end{abstract}
\newpage

\section{Introduction}

Consider long flexible polymers in a not too good solvent. At high 
temperatures, they will form extended random coil configurations which 
can be modeled 
by self avoiding random walks. When $T$ is lowered, there will be a 
critical temperature $T_c$ where the chains start to coagulate and 
the liquid unmixes. This $T_c$ increases with chain length $N$. At the 
{\it theta temperature} defined as $T_\Theta = \lim_{N\to\infty} T_c(N)$ 
also a single (but infinitely long) chain will collapse. Qualitatively, 
this unmixing is described by the mean field theory developed by 
Flory and Huggins \cite{flory}. The phase diagram is sketched in 
fig.1.
\begin{figure}[b]
\begin{center}
\epsfig{file=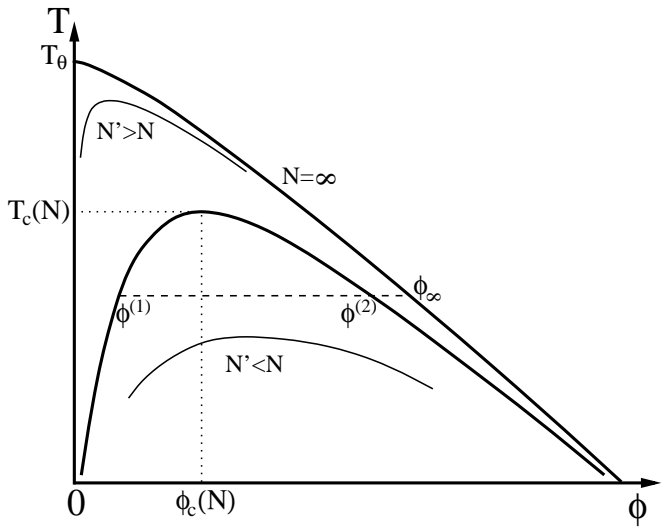,scale=0.65}
\end{center}
\caption{\small Schematic phase diagram for semi-dilute solutions of 
chain polymers. The uppermost curve gives the monomer concentration 
inside an infinitely large collapsed globule under zero outside 
pressure. The lower curves are coexistence curves for fixed chain 
length $N$. These curves are strictly monotonically ordered, 
with the coexistence curve for $N$ below that for $N'$ if $N<N'$.}
\end{figure}

For any finite $N$ we expect that the internal structure of the dissolved 
``particles" (=chains) becomes irrelevant in the infrared limit, i.e. 
close to $T_c(N)$. Thus,
the critical behavior should be fully described by the Ising model, 
i.e. by the $O(n)$ sigma model with $n=0$ \cite{degennes,widom}.
This is indeed supported by all available evidence, but it tells 
only half of the story. As in any critical phenomenon 
there are universal and non-universal properties. While the 
(Ising-)universal aspects (critical exponents and scaling functions) 
must be independent of $N$, all non-universal parameters should 
depend on $N$ systematically, and can be expected to display another 
type of scaling in the limit $N\to\infty$. Also,  
the critical region must become 
smaller and smaller as $N$ is increased, and there must be a cross-over 
to the critical behavior holding at the $\Theta$ point. The latter is 
formally described by a tricritical point in the $O(n)$ model with $n=0$.
As any tricritical point in $O(n)$ models, its upper critical dimension 
is $d=3$, whence it should display mean field behavior with logarithmic 
corrections \cite{degennes}. 

The cross-over should be described by scaling laws which are accessible 
to experiment, but which cannot yet be obtained fully from 
the field theoretic renormalization group. Thus even the basic 
critical exponents are not known, and are the subject of controversial 
speculations. The main theoretical difficulty is that the Ising 
transition and the $\Theta$-collapse have different upper critical 
dimensions, making an $\epsilon$-expansion non-trivial.

Let us denote the monomer density by $\phi$. 
In Flory-Huggins theory one assumes that the entropy per unit volume is 
given by \cite{degennes} 
\be
   S = -{\phi\over N} \ln {\phi\over N} - (1-\phi) \ln (1-\phi) \;,
\ee
and the energy is a quadratic function of $\phi$ independent of $N$. 
Keeping only the mixing entropy, expanding $\ln (1-\phi)$ in 
powers of $\phi$, and dropping the irrelevant linear term in $\phi$, 
the free energy per unit volume is thus given by 
\be 
  \beta F = {\phi\over N} \ln\phi +{1\over 2} v \phi^2 + 
      {1\over 6} w \phi^3 + \ldots \;.    \label{mean-free}
\ee
The $\Theta$ point corresponds to a vanishing of $v$. We assume 
that $v$ is a linear function of $T$, while $w>0$ is constant.
All (mean field) scaling laws can be obtained from this ansatz. 
In particular, the critical density and the distance from the 
$\Theta$ point both turn out to scale as $N^{-1/2}$,
\be
   \phi_c \sim 1/\sqrt{N}   \label{phic}
\ee
and 
\be
   T_\Theta-T_c(N) \sim 1/\sqrt{N}.  \label{Tc}
\ee
For $N=\infty$, the density at $T<T_\Theta$ scales linearly as 
\be
   \phi_\infty \sim T_\Theta-T.    \label{rho}
\ee

Unfortunately, the theory -- being mean field -- predicts also 
the classical value $1/2$ for the order parameter exponent $\beta$. 
For finite $N$, the density difference along the coexistence curve 
(the `order parameter') is predicted to satisfy a scaling law
\be
   \phi^{(2)} - \phi^{(1)} = N^{-1/2} f((T_c-T)\sqrt{N})
                     \label{dphi}
\ee
with $f(x)\sim \sqrt{x}$ for $x\to 0$, and $f(x)\sim x$ for 
$x\to\infty$. A minimal modification which gives the correct value 
of $\beta$ ($\approx 0.325$) was proposed by de Gennes \cite{degennes}. 
He suggested to keep eqs.(\ref{phic}) to (\ref{dphi}), and to adopt 
simply a different behavior for $f(x)$:
\be
   f(x) \sim x^\beta\;,\qquad x\to 0 \;.   \label{fxbeta}
\ee

This ansatz seems theoretically consistent although it cannot be 
derived from an underlying microscopic theory. 
But unfortunately 
eq.(\ref{phic}) is in serious conflict with experiment. Most 
experiments \cite{perzynski,dobashi,shinozaki,chu,xia} and 
subsequent analyses \cite{izumi,sanchez1,sanchez2,enders} agree 
that 
\be
   \phi_c \sim N^{-x_2}      \label{phic2}
\ee
with 
\be
   x_2 = 0.38 \pm 0.01 \;.     \label{x2}
\ee

This discrepancy has given rise to a number of speculations and 
more or less well founded conjectures 
\cite{sanchez1,sanchez2,muthu,stepan,kholod,chera,huillier}. 
One conjecture is that the chains might be partly collapsed at 
the critical point. Since $T_c(N)\to T_\Theta$ for $N\to\infty$, 
the natural first assumption is that the end-to-end distance or 
the radius of gyration should scale as $R\sim \sqrt{N}$, i.e. 
the chains should be free. One might question this since actually 
$T_c(N)<T_\Theta$ for any finite $N$. Accordingly, it was 
suggested in \cite{chera} that chains are partly 
collapsed, 
\be
   R_N \sim N^{x_0}      \label{r2}
\ee
with an exponent $x_0 <1/2$. Actually, this is most unlikely. 
On the one hand, the effective collapse temperature for finite $N$ 
-- defined as that $T$ where $R_N$ is the same value as for ideal 
chains -- is below $T_\Theta$ 
\cite{hegger}. On the other hand, chains must be in contact at 
the critical point (otherwise they would not interact) and can 
penetrate each other. Thus 
there is no force which should compress them beyond the ideal 
shape which maximizes entropy. This argument is very similar to 
that which explains why chains in dense melts are basically ideal.

In spite of these doubts, it seems wise to leave the exponents 
eqs.(\ref{phic2}) and (\ref{r2}) open at the present point, and 
to assume similar scaling laws also for the other observables: 
\be
   \phi^{(2)} - \phi^{(1)} \sim (T_c-T)^\beta N^{-x_1}
                     \label{dphi2}
\ee
and
\be
   T_\Theta-T_c(N) \sim N^{-x_3}  \;.  \label{tau}
\ee
Notice that eqs.(\ref{dphi}) and (\ref{fxbeta}) would give $x_1=
(1-\beta)/2\approx 0.34$. 
Experimentally, the exponents are \cite{widom,wmb}
\be
  x_1 \approx 0.23 - 0.34 \;,\qquad 
     x_3 \approx 0.47 - 0.5.          \label{expo}
\ee

In addition, we can generalize eq.(\ref{rho}) to 
\be
   \phi_\infty \sim (T_\Theta-T)^y    \label{rho2}
\ee
with some unknown exponent $y$, although it seems that {\it all}
authors have assumed that $y=1$, with a single exception to 
be discussed below. 

Since the $\Theta$-point is a tricritical point, one should expect 
logarithmic corrections to these scaling laws \cite{dupla,dupl2}. 
To leading order, they are obtained by replacing $v$ and $w$ in 
eq.(\ref{mean-free}) by their renormalized values for large $N$, 
\be 
  v\to const \;(T-T_\Theta)[\ln N]^{-4/11},\qquad w\to const /\ln N.
\ee
Among others this gives \cite{dupla}
\be
   \phi_c \sim { [\ln N]^{1/2} \over \sqrt{N} }
                                       \label{dupl-phic}
\ee
and
\be
    T_\Theta-T_c(N) \sim N^{-1/2} [\log N]^{-3/11}\;,
                                       \label{dupl-tc}
\ee
From this follows $\phi_c \sim (T_\Theta-T_c)[\log(T_\Theta-T)]
^{7/11}$. 

Similar corrections are predicted for the end-to-end distance of single 
chains at the $\Theta$-point, for the specific heat, and for other 
observables \cite{dupl2} (but not for pure Ising properties such 
as $\phi^{(2)}-\phi^{(1)}$). These corrections for single chains
are not (yet) fully seen in 
simulations. More precisely, simulations \cite{hegger,perm} show 
for most observables corrections which are much larger than those 
predicted by theory, with only weak hints that the theoretical 
predictions are correct for extremely large $N$ \cite{perm}. 

One such case where corrections to mean field behavior are very 
large is the dependence of $\phi_\infty$ on $T_\Theta-T$. Based 
on simulations of very long chains, it was shown in \cite{hegger}
that a best fit is given by $y=0.7$.
But using even longer chains, with $N$ up to $10^6$, it was
concluded in \cite{perm} that this is an effect of very large
corrections to scaling, and that $y=1$ is indeed the most likely 
value. Indeed, as shown in \cite{dupla}, one expects 
logarithmic corrections with the same power as for $\phi_c$,
\be
    \phi_\infty \sim (T_\Theta-T) [\log(T_\Theta-T)]^{7/11} .
                                        \label{dupl-phi}
\ee

Let us forget for the moment any logarithmic corrections, and 
assume that the scaling laws (\ref{phic2}) - (\ref{rho2}) hold. 
We should point out that the exponents $x_i$ and $y$ are not
independent. A simple argument gives 
\be
   x_3\;y \leq x_2 \;.    \label{ineq}
\ee
To derive this, one just needs that $\phi_c(N)\leq \phi_\infty$ 
at fixed temperature $T=T_c(N)$. But this follows from the 
fact that phase coexistence regions grow with $N$: if the 
point $(T,\phi)$ is in the phase coexistence region for some 
$N$, it is also in this region for all $N'>N$ (we have not found 
a mathematically rigorous proof for this, but it seems heuristically 
obvious, and it was assumed tacitly by all previous authors). 

Inserting $y=1$ and the numerical values (\ref{x2}) and (\ref{expo}), 
we see that eq.(\ref{ineq}) is violated. It seems that this 
simple observation has been overlooked in all recent literature, 
and it makes several claims obsolete. One either has to admit that 
$y<1$, or that $x_2$ is much closer to its mean field value than 
suggested by experiments, or (which seems the least likely) 
that $x_3$ deviates strongly from its mean field value.

In view of this unclear situation we decided to perform large 
scale simulations using a novel Monte Carlo algorithm, the 
Pruned-Enriched Rosenbluth Method (PERM) \cite{perm}. We refer 
to this reference and to sec.2 for a description of this 
algorithm. It allowed us to simulate very large systems, with 
chains of length up to 2000 and beyond. Indeed, we could have 
gone even 
further as concerns chain length, but we had problems going 
to systems containing more than $\approx 400$ chains, even if 
these chains are short. This is due to the fact that PERM 
performs excellently at the $\Theta$-point, but becomes less 
efficient at temperatures much below $T_\Theta$. The latter is 
needed for simulations of short chains at the critical point.

Thus all our simulations were done on lattices with finite volume 
$V$ in a regime where finite 
size corrections are important. To perform a detailed finite 
size analysis, we used the histogram method as proposed in 
\cite{wilding-bruce,wm}, and applied to polymers in \cite{wmb}.
In this method, one constructs the microcanonical 
distributions\footnote{Strictly spoken, $P(E,n)$ is the 
microcanonical partition sum multiplied by $e^{-\beta (E - 
\mu n)}$, and normalized so that $\sum_{E,n} P(E,n)=1$.}  
(``histogram") $P(E,n)$ where $E$ is the energy and $n$ is the 
number of chains. Ising universality requires that $P(E,n)$ has 
certain scaling properties which can be used to extrapolate to 
the thermodynamic limit. In the Ising model proper (which is just 
the $N=1$ limit of the lattice polymer model studied below), 
$2nN/V-1$ is replaced by the magnetization $M$, and $P(E,M)$ 
is symmetric under $M\to -M$. This symmetry is broken for $N>1$, 
but restored at the critical point (i.e., for $E\approx E_c$ 
and $n\approx n_c$ in the thermodynamic limit. It was claimed 
in \cite{wilding-bruce,wm,wmb} that the dominant symmetry 
breaking term in the vicinity of $(E_c,n_c)$ can be removed by 
an affine transformation 
\be
   \cE = E-rn\;,\qquad \cn = n-sE    \label{mix}
\ee
with $N$-dependent parameters $r$ and $s$ (``field mixing" 
\cite{rehr}).
We found that this is not true in our case. Although parameters 
$r,s$ can be found such that the marginal distributions $P(\cn)$ 
and $P(\cE)$ agree with the Ising universal curves within error 
bars (implying also that $P(\cn)$ is symmetric around $\cn_c$), 
the 2-dimensional distribution does 
not become more symmetric by this transformation. Nevertheless, 
the possibility to compare with the precisely known critical 
magnetization distribution of the Ising helps enormously in 
fixing the critical point, and extracting critical parameters. 
We thus basically verified that the histogram
method gives very reliable (and large!) finite size corrections,
allowing us to obtain precise critical parameters from fairly
small system sizes.

The algorithm and computational details are discussed in the 
next section. In sec.3 we describe the histogram method, and 
our main results are presented in sec.4. One interesting aspect 
of PERM is that it allows to compute free energies with high 
precision. We use this to compare directly with the mean field 
ansatz eq.(\ref{mean-free}). We conclude with a 
discussion of our results in sec.5.

\section{Algorithm and Computational Aspects}

The Pruned-Enriched Rosenbluth Method (PERM) \cite{perm} is 
a chain growth algorithm based on the Rosenbluth-Rosenbluth (RR) 
method \cite{rosen} and on ``enrichment" \cite{wall} or copying of 
successful partial chains. As is well known \cite{batoulis}, the 
main drawback of the RR method is that it leads to weighted samples 
with very uneven weights. This is counterbalanced in PERM by 
enrichment (when a large weight chain is copied $k$ times, all 
$k+1$ copies receive $1/(k+1)$ times the original weight) and 
by {\it pruning}: low weight chains are either pruned (deleted) 
or, with the same probability, doubled in weight. Similar 
algorithms have been used in \cite{garel,velikson}. There, 
however, they were implemented in a `breadth first' way, by 
keeping a large population of chains simultaneously in 
memory. Such a strategy would have been not feasible for the 
very large systems (up to several hundred thousand monomers)
treated in this paper. Instead, we used a recursive `depth 
first' strategy. The basic algorithm is described in detail 
in \cite{perm}, some further tricks to make it faster and more 
robust are discussed in \cite{stiff,fold}. We refer to these 
papers for details. We just mention that we hand-coded the 
recursion without using recursive function calls, in contrast 
to the algorithm shown in the appendix of \cite{perm}, since 
we otherwise would have had problems with storage. All simulations 
reported in this paper were done on workstations and used 
less than 50 MB main memory.

As designed originally, PERM applies to classical (distinguishable) 
particles which are tied together to form a chain polymer. 
To simulate systems with several polymer chains which are 
indistinguishable and not connected, we need some modifications. 
The first is that each time when we start a new chain, we have 
to chose the location of its first monomer anywhere in the not 
yet occupied volume. For the $k$-th chain in volume $V$, this 
implies a Rosenbluth factor $V-(k-1)N$. The total Rosenbluth 
weight for a system consisting of $n$ chains is then 
$\prod_{k\leq n}(V-(k-1)N)$ times the product of 
proper Rosenbluth factors for the second, third, ... $N$-th
monomers in the chains. Secondly, we have to multiply the 
final weight by $1/n!$ since there are exactly $n!$ possibilities 
to build up a configuration of $n$ chains sequentially. 

In PERM one uses the product of Rosenbluth and Boltzmann weights
of partially assembled systems to steer 
doubling and pruning. In order to have a smooth dependence of the 
doubling and pruning thresholds on the number of already assembled 
monomers, we omitted the above factors during the growth, and 
added them later when the system was already built up.

All simulations were done on the simple cubic lattice with 
helical boundary conditions (lattice sites are indexed by a 
single integer $i$, and $i+V \equiv i$). Chains were modeled by 
self-avoiding walks with attractive energy $\epsilon = -1/k_B$ 
between each pair of neighboring non-bonded monomers.
Chain lengths were powers 
of 2. Volumes were also powers of two, $V=2^m$ with 
$m$ such that systems at the critical point had roughly 100 chains.
Thus $V$ changed from $2^{12}$ for $N=8$ to $2^{21}$ for $N=2048$. 
CPU times (on DEC Alpha machines with 400 Mhz) ranged from a few 
hours for $N=8$ to roughly two weeks for $N=2048$.

In these simulations we used one constant chemical potential $\mu$ 
for each chain, and another potential $\mu'$ for each monomer. 
Since we measured 
observables only after chains had been finished, this corresponds 
to an overall fugacity $z=e^\mu e^{N\mu'}$ per chain. We measured: 
\begin{itemize}
\item The canonical partition sum $Z_n$ for each $n=1,2,\ldots, 
n_{\rm max}$. This is needed anyhow for the control of doubling and 
pruning. 
\item The average energy $E_n$ of the total configuration containing 
$n$ chains, $n=1,2,\ldots, n_{\rm max}$.
\item The average energy $e_n$ of the $n$-th inserted chain at the 
moment of its insertion. Notice that $E_n$ is approximately equal 
to $\sum_{k=1}^n e_k$, but not exactly.  
\item The average r.m.s. end-to-end distance $R_{N,n}$, averaged over 
all chains. 
\item The end-to-end distance $r_{N,n}$ for the last inserted chain. 
Again, $R_{N,n}$ is approximately but not exactly 
equal to $r_{N,n}$.
\item A histogram which contains for each pair $(E,n)$ the 
microcanonical partition sum multiplied by $e^{-\beta E}z^n$.
\end{itemize}

We do not present results for chains shorter than $N=8$, mainly 
because this requires simulations at rather low temperatures. The
algorithm becomes inefficient there. For instance, for the 
Ising model ($N=1$) we had no problems to simulate lattices of 
size $4^3$, verifying thereby that the algorithm works in 
principle also on this extreme case. But already for systems 
with $V=8^3$ we encountered problems. This is not too surprising. 
In spin language, we start simulations with all spins down. We 
chose random sites where spin is still down and flip it up, keeping 
track of the Boltzmann weights by giving each configuration a 
weight different from 1. If this weight becomes too large or too
small, we copy or prune, respectively. Finally, if pruning did not 
kill us, we end up at the configuration with all spins up. The 
total weight of this state should be the same as that of the 
starting configurations with all spins down. While this symmetry 
is exactly respected by the Rosenbluth method without pruning 
or copying, it is not by PERM due to the stochastic nature of 
pruning. In the long run, of course, the symmetry will be approached 
closer and closer. But this is due to very rare events with very 
large weights. It not only involves huge statistical errors but 
also a systematical bias unless one is very careful and does not 
estimate statistical errors from fluctuations within small samples.
These problems have to be kept in mind when dealing with $N>1$. 
In this case there is no symmetry to check, but there is the same 
danger of underestimating the contribution of rare outlyers. We 
hope to have minimized this danger by using very large samples, 
typically $10^6$ to $10^7$ independent ``tours" in the terminology 
of \cite{perm}.

\section{Histogram Method and Finite Size Scaling}

Let us consider for the moment an Ising system in $d$ dimensions 
with lattice size $L$. We denote by ${\cal Z}_L(T,h)$ the canonical 
partition sum, and by $Z_L(E,M)$ the number of states with energy 
$E$ and magnetization density $M$. The {\it histogram} $P_L(E,M)$ 
is defined as 
\be
   P_L(E,M) = {e^{-\beta E + h L^dM}\over {\cal Z}_L(T,h)}
       Z_L(E,M)\;.
\ee
For vanishing external magnetic field $h$ and at $T=T_c$ it should 
scale as \cite{wilding-bruce,wm}
\be
   P_L(E,M) \approx L^{\beta/\nu-1/\nu} g(L^{\beta/\nu}M,
              (E-E_c)L^{-1/\nu})            \label{pem}
\ee
where $E_c$ is the average energy at the critical 
point, and $\beta = 0.327\pm 0.001$ and $\nu = 0.630 \pm 0.001$ 
\cite{bloete,talapov} are the usual critical indices. The function 
$g$ is universal up to rescalings of its value and arguments by 
arbitrary factors. Summing over $E$ resp. $M$ we obtain the 
distributions of magnetization resp. energy, 
\be
   P_L(M) \approx L^{\beta/\nu} \tilde{p}_M(L^{\beta/\nu}M)  \label{pm}
\ee
and
\be
   P_L(E) \approx L^{-1/\nu} \tilde{p}_E((E-E_c)L^{-1/\nu}) \;.    \label{pe}
\ee
The energy scaling function $\tilde{p}_E(x)$ has a single 
maximum. In contrast, the magnetic scaling function 
$\tilde{p}_M(x)$ has two (symmetric) maxima for $d<4$. 

This is 
the main point which a finite size scaling analysis has to take into 
account correctly. Naively, the critical point is defined as that 
temperature where $\tilde{p}_M(x)$ changes from being 
single-humped to double-humped. But strictly, this is true only 
in the thermodynamic limit, and analyses which neglect this when 
estimating $T_c$ \cite{madden,mackie,yan} can have very large 
systematic errors.
More precisely, in $d=3$ the ratio $\tilde{p}_M(0)/\max_x
\tilde{p}_M(x)$ is roughly 0.44, with an error probably $<0.02$ 
\cite{wm}. In the $(E,M)$-plane, the
distribution $P_L(E,M)$ at $h=t=0$ has two peaks located on the 
two halves of a {\sf U}-shaped support.

Let us now consider a system like a critical gas where $M$ is 
replaced by the particle density $\rho$ ($=\phi/N$), and the 
symmetry $M\to-M$ is lost.
In the following we shall use the total particle number $n$ 
instead of $\rho$.
According to \cite{wilding-bruce,wm}, one just has to replace 
$E$ and $n$ by linear combinations
\be
   \cE = E - r\;n                \label{cu}
\ee
and 
\be
   \cn = n - s\;E              \label{cm}
\ee
with suitable constants $r$ and $s$. One arrives at scaling 
laws for $\cE$ and $\cn$ which formally coincide exactly with 
eqs.(\ref{pem}-\ref{pe}) except for the fact that also $\cn_c=
\langle\cn\rangle$ is different from zero and that $\cn-\cn_c$, 
being an extensive quantity, replaces not $M$ but $L^dM$:
\be
   P_L(\cE,\cn) \approx L^{\beta/\nu-1/\nu-d} g((\cn-\cn_c)
    L^{\beta/\nu-d}, (\cE-\cE_c)L^{-1/\nu}) \;.  \label{P-transf}
\ee
This conjecture was tested 
in \cite{wilding-bruce,wm,wmb} mainly by verifying that the 
projected 1-dimensional distributions $\tilde{p}_L(\cn)$ and 
$\tilde{p}_E(\cE)$ agreed numerically with the Ising scaling 
functions.
For polymer solutions we expect $r$ and $s$ to scale with $N$ in 
a universal way. 

Notice that the justifications for eqs.(\ref{cu}) and (\ref{cm}) are rather
different. Replacing the energy density $E$ by a linear combination $E -
r\,n$ just corresponds to changing the internal energy of the particles.
Since that is arbitrary anyhow, we see that eq.(\ref{cu}) is very natural.
It just corresponds to a shift in the internal energies such that both
phases have the same energy density per unit volume. No such interpretation
can be given for eq.(\ref{cm}).

Let us assume that we have made the simulations at a temperature near
$T_c$.  We first obtain rough estimates of the critical fugacity $z_c$ and
of $r$ by demanding that both peaks in $P_L(n)$ have the same height, and
that the two peaks in $P_L(\cE,n)$ occur at the same value of $\cE$. A
typical result, for $N=128$, is shown in fig.2a. By summing over all $E$,
we obtain the chain number distribution shown in fig.3a. Notice the very
asymmetric shape of both distributions. Notice also that the ratio between
the height of the central minimum of $P_L(n)$ and its maximal value is not
exactly 0.44, indicating that we have not simulated exactly at $T_c$.

\begin{figure}[t]
\begin{center}
\vglue -8mm
\begin{minipage}[t]{5.1cm}
\caption{\small (a) Contour plot of $P_L(\cE,n)$ for chains of length 
$N=256$ in a lattice with size $L=64$. Notice the asymmetric shape. 
For the Ising model, the contour plots would be symmetric under 
reflection on the vertical axis. The normalization is arbitrary.
(b) Same data as in panel a, but after transforming $n\to \cn$ by means 
of eq.(\ref{cm})
and readjusting the other parameters. The parameter $s$ is $0.003$.
(c) Same data again, but using the nonlinear transformation 
eq.(\ref{mix-nonlin}).}
\end{minipage}
\begin{minipage}[t]{10.0cm}
\psfig{file=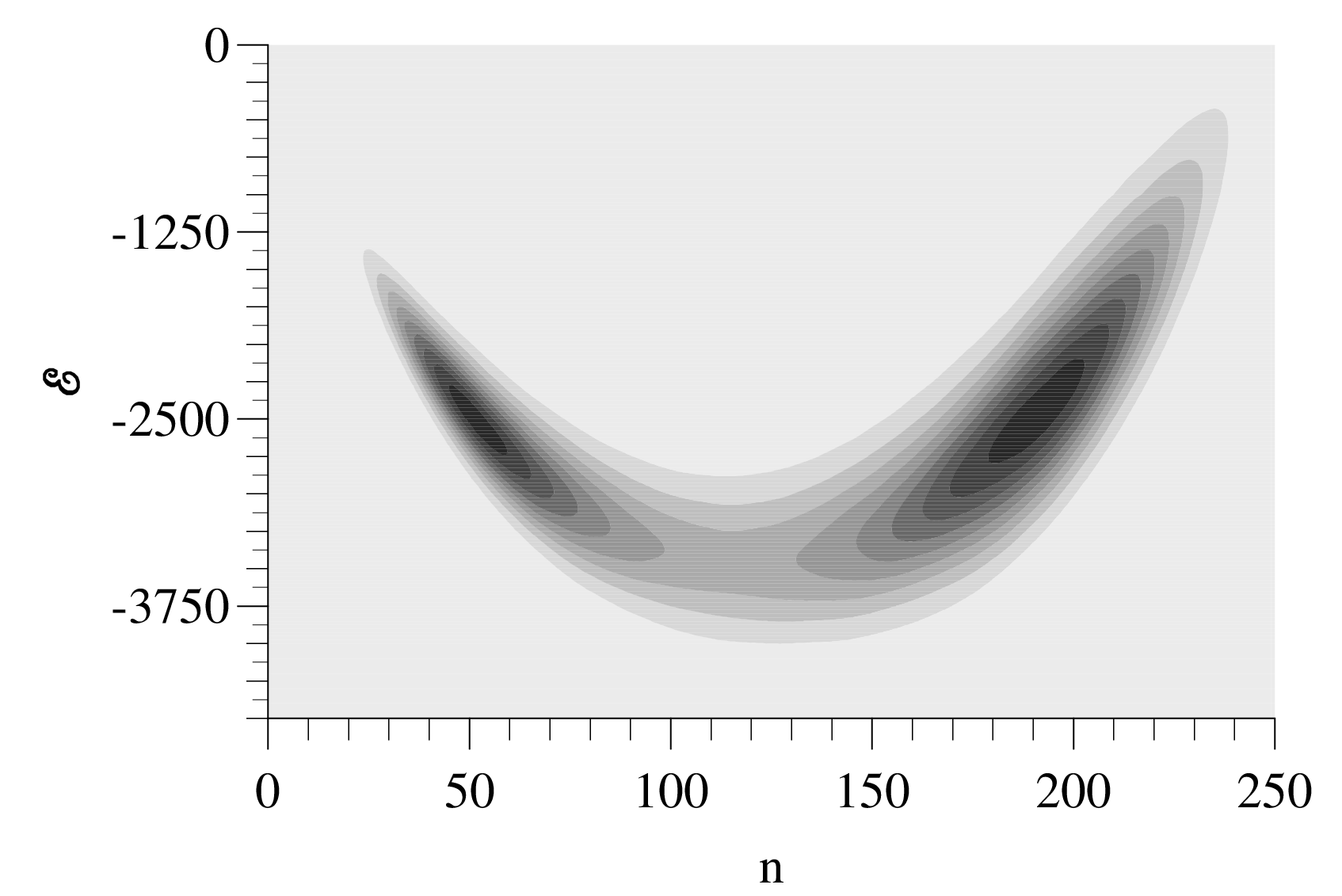,scale=0.36,angle=270}
\psfig{file=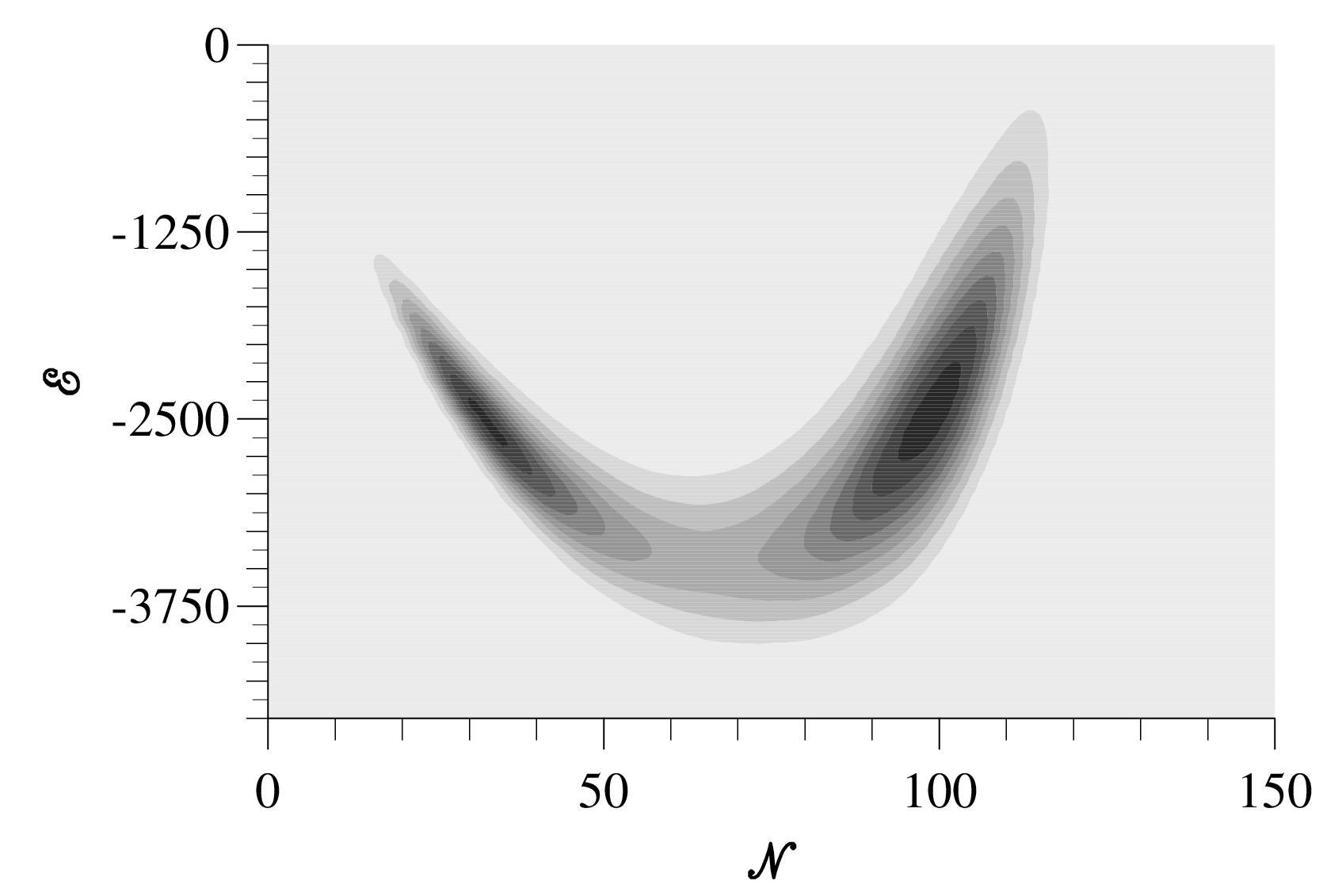,scale=0.36,angle=270}
\psfig{file=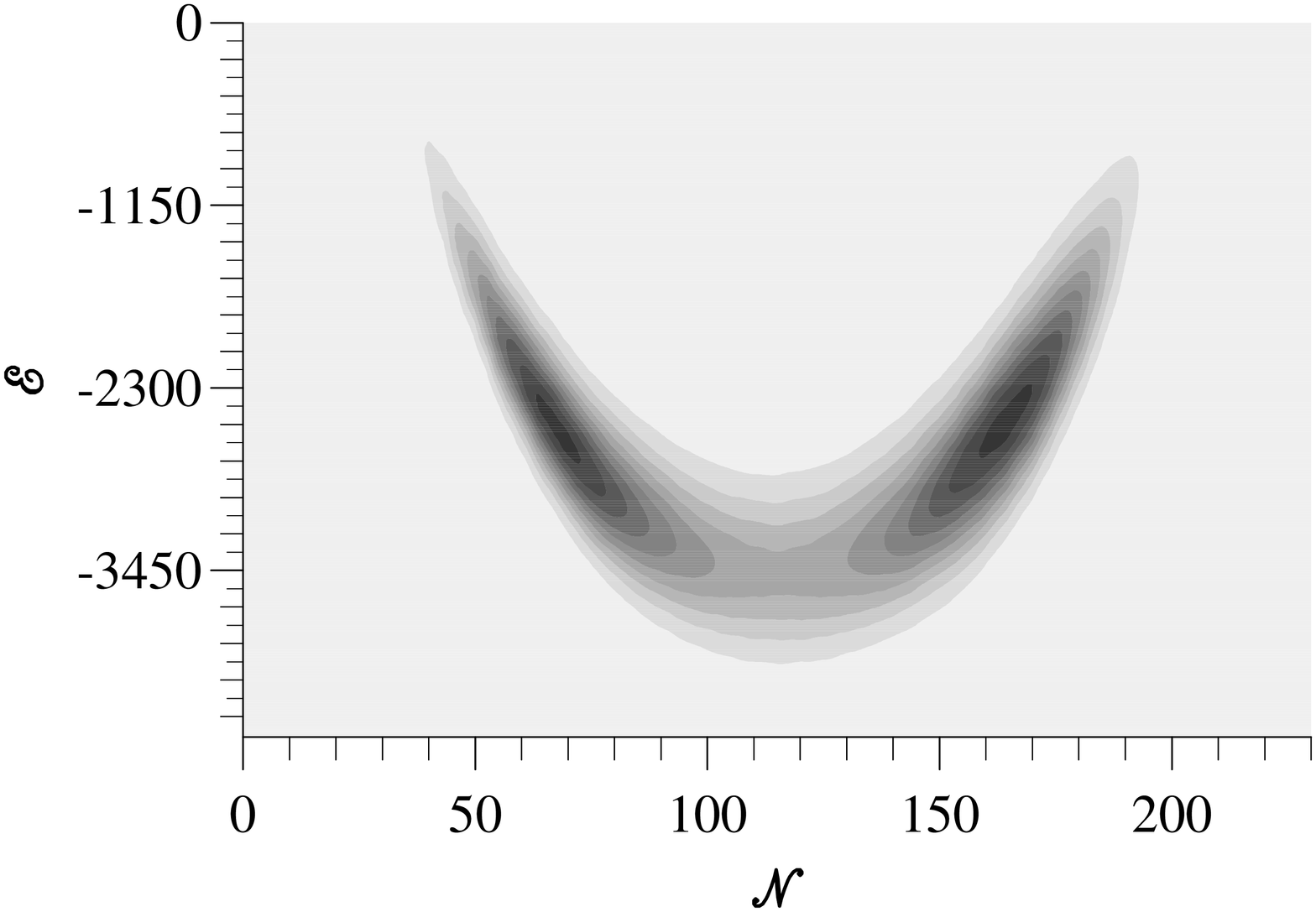,scale=0.36,angle=270}
\end{minipage}
\end{center}
\vglue -8mm
\end{figure}

\begin{figure}[t]
\begin{center}
\vglue -8mm
\begin{minipage}[bh]{10.0cm}
\epsfig{file=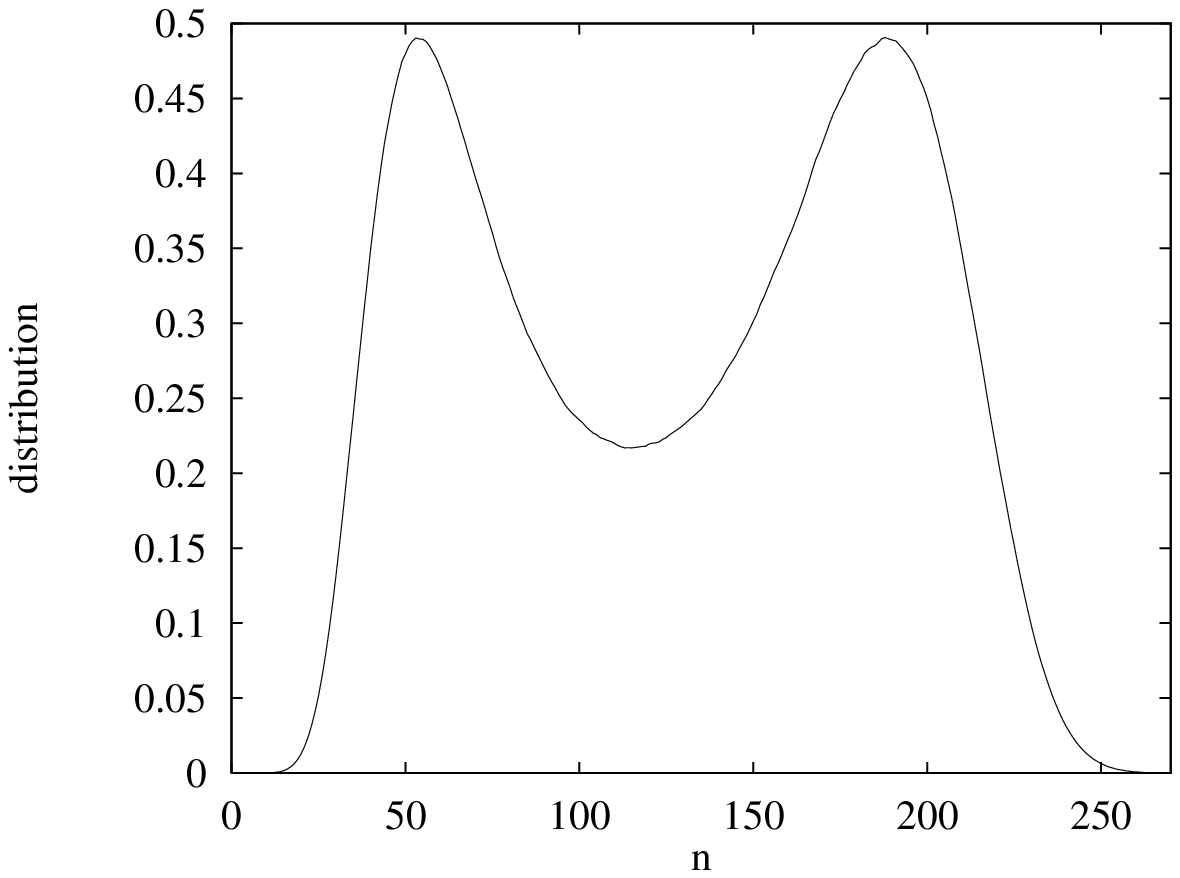,scale=0.8}
\epsfig{file=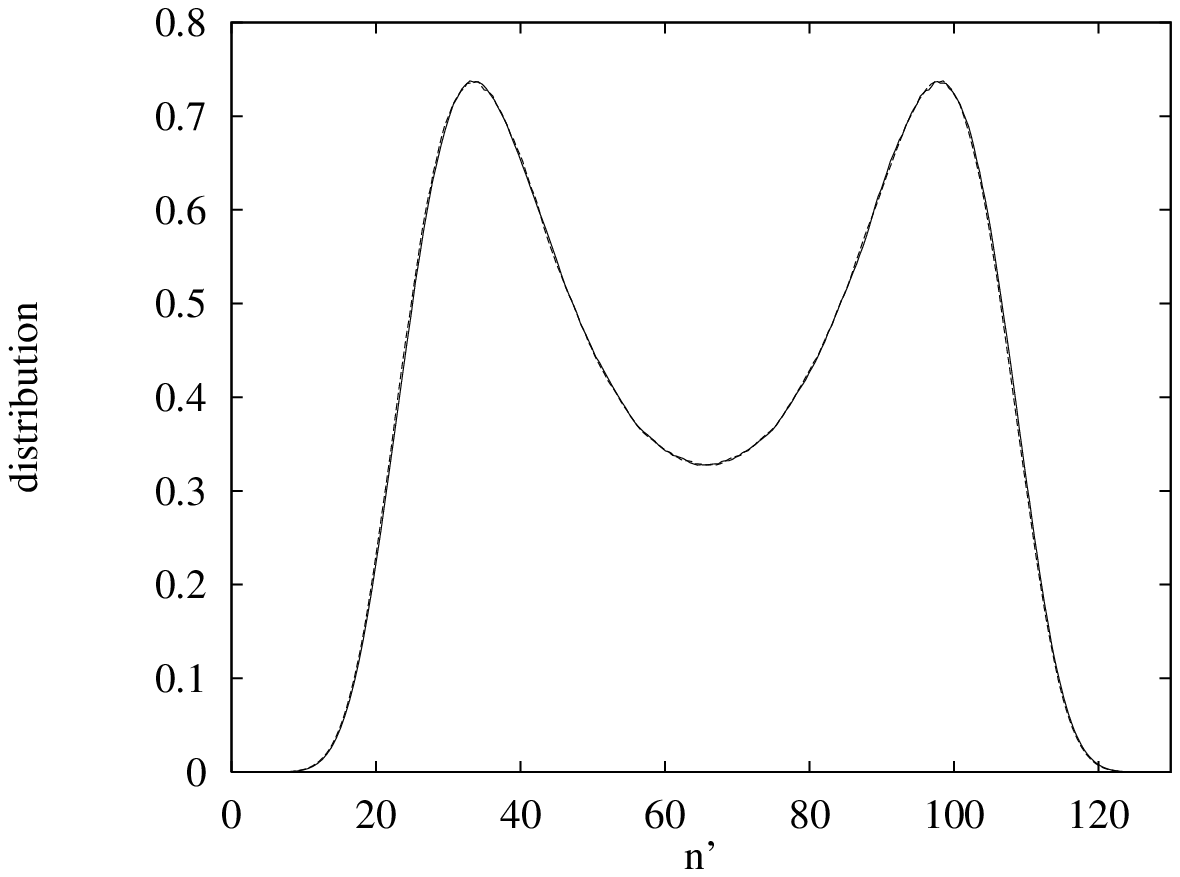,scale=0.8}
\end{minipage}
\end{center}
\vglue -5mm
\caption{\small (a) Distribution $P_L(n) \equiv z^nZ_n$ (up to an arbitrary 
normalization factor) against $n$, obtained 
by projecting the density of fig.2 onto the horizontal axis. 
(b) Same as panel a, but after transforming $n\to \cn$. 
To demonstrate the symmetry of the distribution, both curves 
$P_L(\cn)$ and $P_L(2\cn_c-\cn)$ are superimposed. As in the 
previous figures, the normalization is arbitrary.}
\end{figure}
In a first refinement we now determine a crude estimate of $T_c$ by
re-weighting $P_L(\cE,n)$ such that the minimum in the projected
distribution $P_L(n)$ has a height equal to 0.44 times the average value
of the maxima. We then fit $s$ such that $P_L(\cn)$ becomes symmetric under
$\cn-\cn_c \to -(\cn-\cn_c)$, where $\cn_c$ is the minimum position, and
readjust at the same time the fugacity $z_c$ such that both peaks have
again the same height.  The fact that we always do find a value of $s$ such
that $P_L(\cn)$ becomes approximately symmetric and equal to the Ising
scaling function is highly non-trivial.

The last readjustment will in general have shifted the peaks in the
$(\cE,\cn)$-plane such that they occur no longer at the same value of
$\cE$. We thus repeat the fit of $r$ and estimate more precise values of
$T_c,z_c$, and $s$ by repeating the second step. In principle, we could
iterate this procedure until we reach convergence, but already after the
second refinement the two peaks of $P_L(\cE,\cn)$ will occur at the same
$\cE$. We then check that $P_L(\cn)$ and $P_L(\cE)$ agree with the scaling
functions given in \cite{wm} within the statistical errors. The final
results for the raw data shown in figs.2a and 3a are shown in figs. 2b and
3b. For other chain lengths, results are very similar.

We thus have succeeded in finding field mixings such that the 1-d
projections of the 2-d histogram agree with the Ising case.  Exactly this
was done also in \cite{wilding-bruce,wm,wmb}. But we see from fig.2b that
the symmetry of the projected density $P_L(\cn)$ is misleading. The 2-d
histogram has not become more symmetric by replacing $n$ by $\cn$. We have
not tried to apply formal measures of symmetry to fig.2, but it seems that
fig.2b is rather less symmetric than fig.2a.

We see thus that linear field mixing is questionable. We tried 
a number of alternatives. By far the most successful numerically 
is the following. We first subtracted from $E$ a term linear in $n$ 
(corresponding just to a redefinition of internal energy), and then 
applied a conformal transformation with a scale factor which depends 
only on $n$. We did not try to optimize systematically since this 
is heuristic anyway, but good results were obtained with power law 
factors,
\be
   \cn = (n/\langle n\rangle)^\alpha n \;,\qquad 
   \cE = (n/\langle n\rangle)^\alpha (E-r\,n)\;.
                         \label{mix-nonlin}
\ee
The same data shown already in figs.2a and 2b are plotted 
against these variables in fig.2c, with $\alpha =- 0.3$. We see 
indeed a much more symmetric distribution. One-dimensional 
projections from this distribution are not very different from
those obtained with linear mixing, and are not shown.
\begin{figure}[hb]
\begin{center}
\vglue -8mm
\begin{minipage}[t]{10.0cm}
\epsfig{file=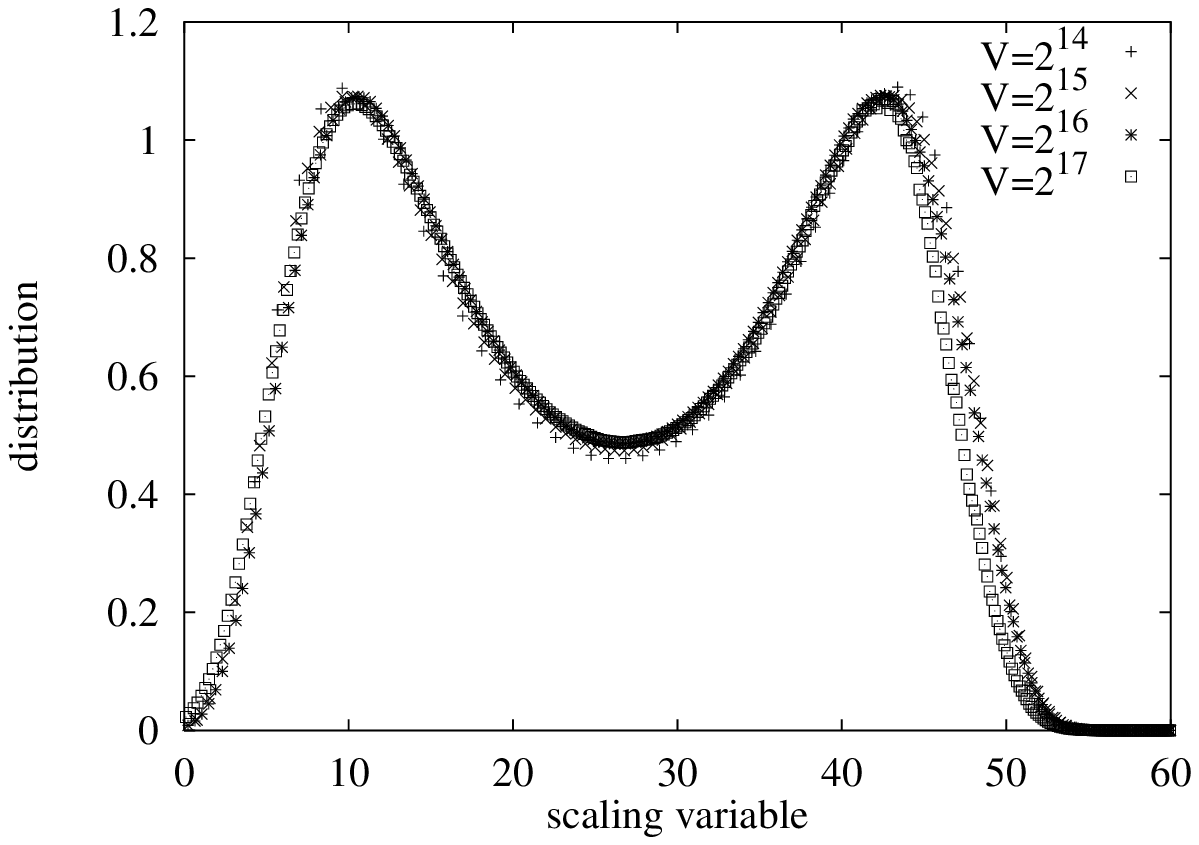,scale=0.8}
\epsfig{file=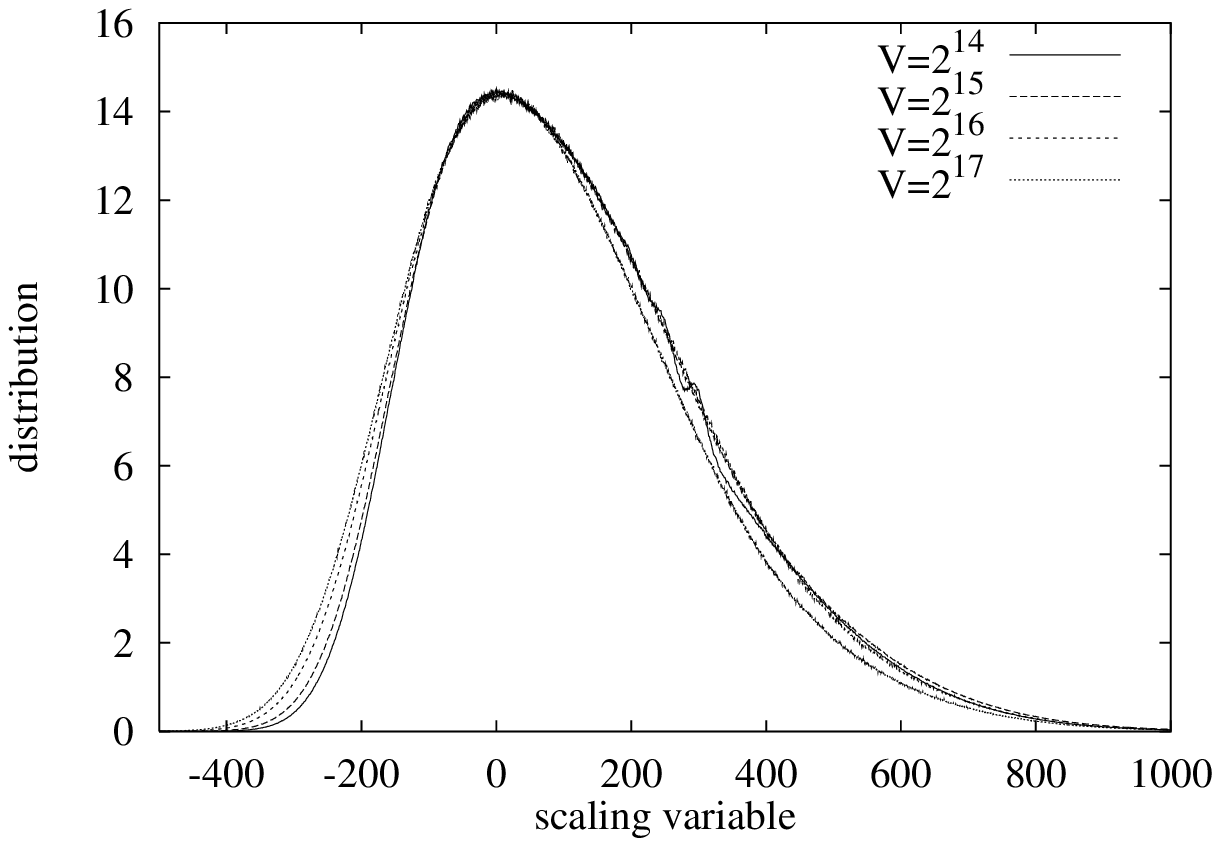,scale=0.8}
\end{minipage}
\end{center}
\vglue -5mm
\caption{\small (a) Scaling plot of the distribution of transformed chain 
numbers $\cn$ for $N=128$ and $T=3.168$. Within statistical errors, 
the latter is our estimate for $T_c$. The critical monomer density
was assume to be $\phi_c = 0.1544$. (b) The analogous 
distribution of the transformed energy $\cE$.}
\end{figure}

These different mixing ansatzes showed us that the estimates of $T_c$,
$n_c$ and of the critical fugacity are very robust. Heuristically, this can
be explained by the fact that the support of $P_L(E,n)$ is a very narrow
band, i.e. the variance of $E$ for fixed $n$ is rather small. We thus
conclude somewhat paradoxically that we cannot verify a basic assumption of
\cite{wilding-bruce,wm,wmb}, but we agree that the histogram method is
extremely helpful for extracting critical parameters.

To verify the correctness of the approach (and of the simulations!), we
checked the scaling with $L$ for one chain length, $N=128$. According to
eq.(\ref{pm}), the values of $L^{-\beta/\nu}P_L(\cn)$ should coincide when
plotted against $L^{\beta/\nu}\cn$. This scaling plot is shown in fig.4a,
while the analogous scaling plot for the energy distribution is shown in
fig.4b. We see perfect agreement. For other chain lengths we usually
simulated at two lattice sizes in order to check for obvious
inconsistencies, but we did not check systematically the finite size
scaling.

\section{Results}

\begin{figure}[b]
\begin{center}
\vglue -8mm
\psfig{file=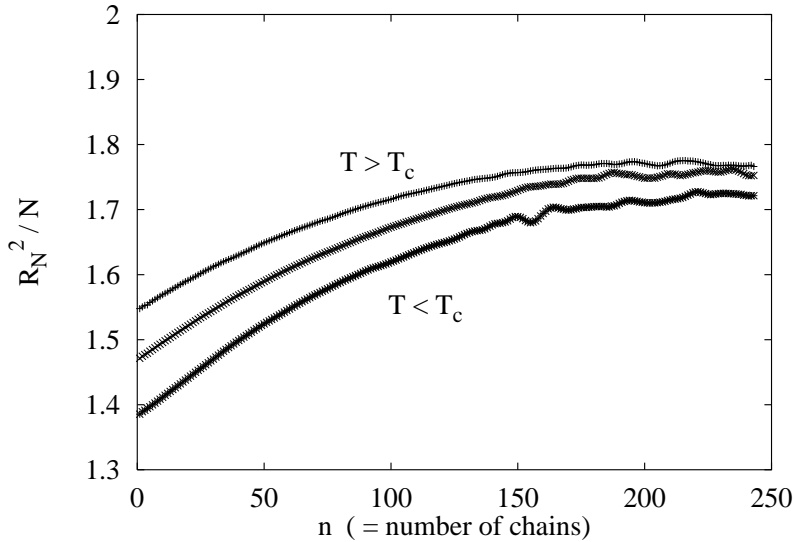,scale=0.85}
\end{center}
\vglue -5mm
\caption{\small Swelling factors $R_N^2/N$ for fixed $N=128$ at three 
temperatures, $T=3.03, 3.1655, $ and $3.30$ (bottom to top), plotted 
against the number of chains $n$. The central value is close to the 
estimated $T_c$. All data were obtained with $L^3 = 2^{16}$ sites.}
\end{figure}
As a first result we show the swelling factor $R_N^2/N$ for three 
different temperatures as a function of the number $n$ of chains 
(fig.5). These results were obtained at fixed $L$ and $N$. The 
critical point corresponds to the central curve and to $n_c \approx 
79$. From this we can make several interesting observations:\\
1) All chains are swollen, i.e. $R_N>\sqrt{N}$ in all cases.\\
2) The swelling is weakest for $n=1$, i.e. for isolated chains, 
and increases as the chain density is increased. For large $n$ it 
seems to saturate at a value $<2$. This agrees with 
the fact that all three temperatures are below the $\Theta$ point, 
and chains at $T_\Theta$ are slightly swollen, with $R_N^2/N\to 1.7 
- 2$ for $N\to\infty$ \cite{hegger,perm}. For large $n$ the chains 
feel mainly the slight imbalance between self avoidance and 
attraction for short chains which makes them swollen like very 
long $\Theta$ chains, but for small $n$ the attraction from the 
other chains is missing and the chains are slightly less swollen. \\
3) There is rather weak dependence on $T$, and this dependence 
weakens further as $n$ increases.
\begin{figure}[ht]
\begin{center}
\vglue -3mm
\psfig{file=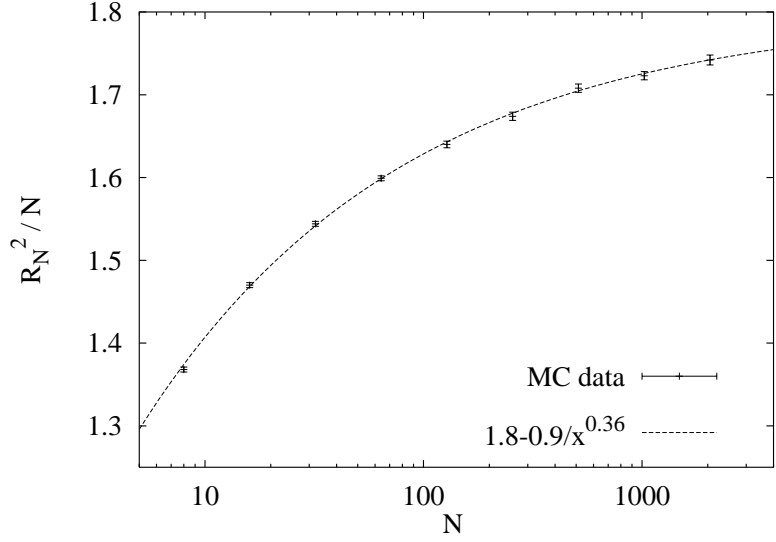,scale=0.85}
\end{center}
\vglue -5mm
\caption{\small Swelling factors at the critical point against $N$. The 
dashed line is a fit with the function $1.8-0.9x^{-0.36}$. This fit
has no particular significance except for the fact that the limit 
$N\to\infty$ agrees with the swelling of infinitely long $\Theta$ 
polymers.}
\end{figure}

As a result of the last point, the systematic uncertainty in 
$R_N^2/N$ due to the uncertainty of $T_c$ is negligible. Not 
negligible, in contrast, is the dependence on the estimate of 
$n_c$. Indeed, due to the very small statistical errors in fig.5, 
the uncertainty of $n_c$ is the largest source of error for $R_N$.

Results for $R_N^2/N$ at the critical point obtained 
from fig.5 and from similar figures for other values of $N$ are 
shown in fig.6. We see that the swelling increases slightly 
with $N$, as also found in \cite{wmb}. But in contrast to these 
authors we see that the swelling saturates, ruling out a growth 
of $R_N$ with a power $>1/2$. Of course, such a growth would seem 
very strange and was rejected on this ground in \cite{wmb}. But 
we can even extrapolate to $N=\infty$ (dashed curve in fig.6), 
and verify that we obtain in this limit the same swelling as for 
isolated $\Theta$ polymers.

\begin{figure}[ht]
\begin{center}
\vglue -3mm
\psfig{file=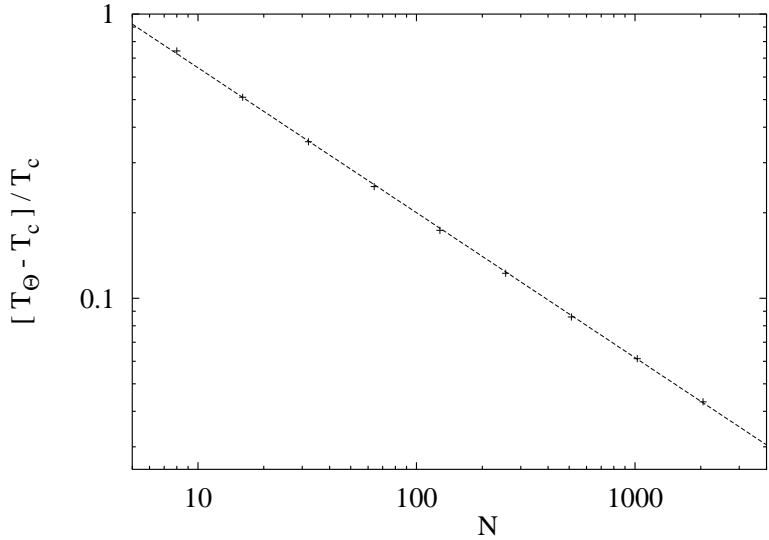,scale=0.85}
\end{center}
\vglue -5mm
\caption{\small Log-log plot of $(T_\Theta - T_c(N))/T_c(N) $ 
against $N$ (crosses). The dashed line has slope $-0.51$.}
\end{figure}
To estimate the exponent $x_1$ directly, one would have to simulate 
much larger systems than we could afford. But one can obtain this
exponent indirectly from finite-size scaling exactly at $T_c$. Let 
us denote by $\Delta \cE$ and $\Delta \cn$ the rms widths of 
$p_L(\cE)$ and $p_L(\cn)$, respectively. Keeping the $N$-dependent
non-universal factors omitted in eq.(\ref{P-transf}), one easily
shows that
\be
   \phi^{(2)}-\phi^{(1)} \sim (T_c-T)^\beta \, {N\over L^d}
            \Delta \cn [\Delta \cE]^\beta.     \label{dphi3}
\ee
We tried to use this for estimating $x_1$. But the results were 
rather disappointing, presumably because of the uncertainty in the 
field mixing. In the linear mixing ansatz, eqs.(\ref{cu},\ref{cm}) 
were actually written in \cite{wilding-bruce,wmb} as a map
$(\cE,\cn) = (1-rs)^{-1}(E - rn,n-sE)$. We have omitted the factor 
$(1-rs)^{-1}$ up to now since it is irrelevant for everything 
we did up to now, but it does become relevant in eq.(\ref{dphi3}).
But we also had argued that eq.(\ref{cu}) is very natural as it is 
written, without the factor $(1-rs)^{-1}$. Thus we propose that 
the correct linear mixing for our purpose is 
\be
   \cE = E - r\;n,\qquad \cn = (1-rs)^{-1} (n - s\;E) = n-{s\over 
     1-rs} \cE .
\ee
Using this, we obtain $x_1 \approx 0.18$, albeit with large error
bars which are hard to estimate since they are a mixture of 
statistical and systematic errors. 

Alternatively, we also tried 
the nonlinear transformation eq.(\ref{mix-nonlin}), and we 
tried to use the untransformed width $\Delta n$ instead of 
$\Delta \cn$. Although the detailed numbers entering into 
eq.(\ref{dphi3}) are then quite different from those 
obtained with linear mixing, the resulting values of $x_1$ 
were again $\approx 0.15$ to 0.2, confirming thus the estimate 
$x_1 \approx 0.18$. In view of its 
uncertainty, this estimate is presumably compatible with the lower 
end $x_1\approx 0.23$ of phenomenological analyses, but it seems 
very hard to reconcile it with the prediction $x_1=(1-\beta)/2 = 
0.34$ of eqs.(\ref{dphi},\ref{fxbeta}).

Estimates of $T_c(N)$ are shown in fig.7. More precisely, this is
a log-log plot showing $(T_\Theta - T_c(N))/T_c(N)$ against 
$N$. We see a very clean scaling law with exponent $x_3=0.51 
\pm 0.01$, suggesting that the mean field value $1/2$ is indeed 
correct. We should however realize that we are not yet very far in 
the scaling regime, in spite of the large values of $N$. Thus, 
using slightly different scaling variables as, e.g., 
$(T_\Theta - T_c(N))/T_\Theta$ instead of the variable used in 
fig.7, might lead to slightly different exponents. Also, this 
plot is very sensitive to the exact value of $T_\Theta$. In order 
to reduce this uncertainty further, we have made additional 
simulations of single very long polymer chains using the same 
routine as in \cite{perm}, but with even larger $N$ (up to 
$1.6\times 10^6$) and even larger system sizes (up to $512^3$ 
sites). Our best estimate is now $T_\Theta = 3.717\pm 0.002$.
This uncertainty contributes less than $0.01$ to the uncertainty 
of $x_3$. Finally, according to eq.(\ref{dupl-tc}) we should see 
weak logarithmic corrections. We believe that we do not see them 
in fig.7 because of the just mentioned uncertainties. 
In summary, we can say that $x_3=1/2$ is the most likely value.

Finally, in fig.8 we show our estimates of the critical density. 
For our largest values of $N$ (which agree roughly with the 
longest chains used in experiments) we see a power law with 
exponent $x_2 \approx 0.38$. This agrees perfectly with experiment
and with phenomenological analyses \cite{dobashi}-\cite{sanchez2}. 
If we accept this as the true critical exponent, and accept at the 
same time the mean field exponent $y=1$, we run into the problem 
posed by inequality (\ref{ineq}). But we see also from fig.8 
that there are very large corrections to scaling. If we define 
effective $N$-dependent exponents by fitting locally, they 
increase slightly but systematically with $N$. 
\begin{figure}[ht]
\begin{center}
\vglue -3mm
\psfig{file=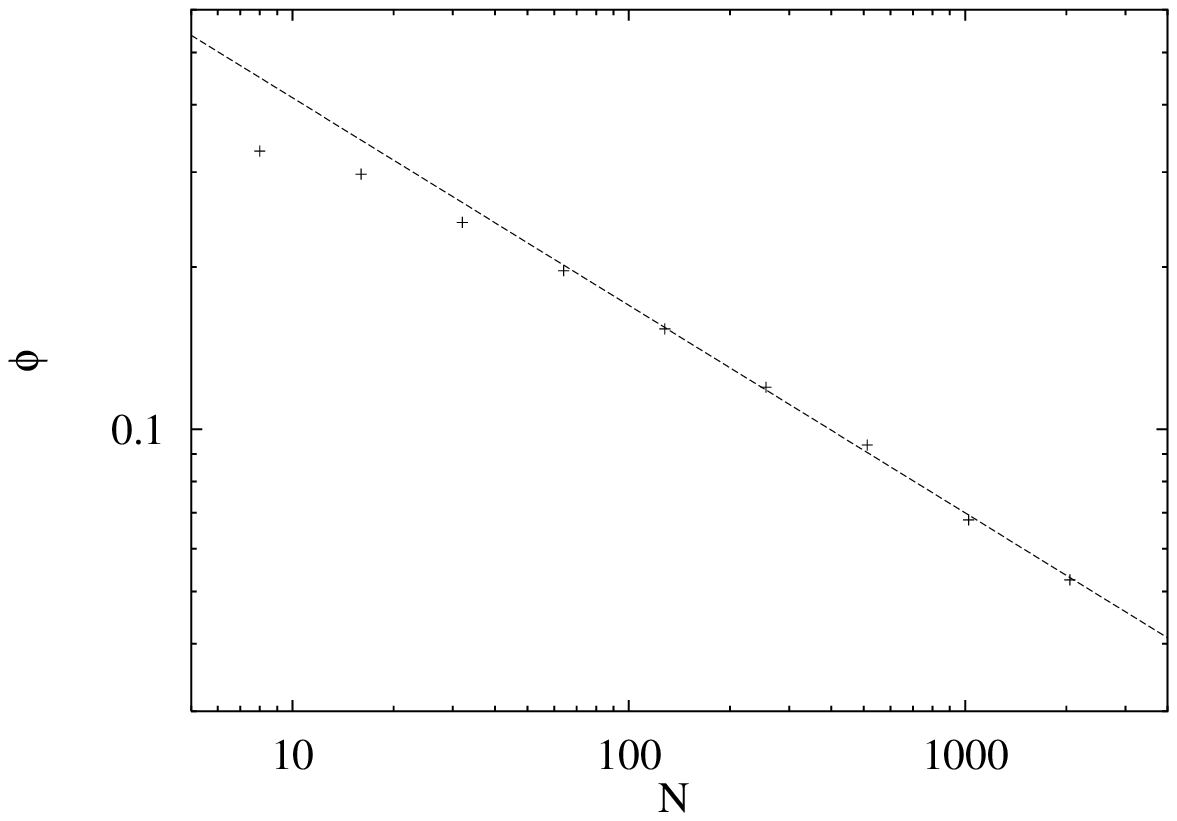,scale=0.85}
\end{center}
\vglue -5mm
\caption{\small Log-log plot of $\phi_c(N)$ against $N$. The dashed 
line has slope $-0.385$. It fits the data for large $N$, but there 
are very substantial deviations at small $N$.}
\begin{center}
\vglue -3mm
\psfig{file=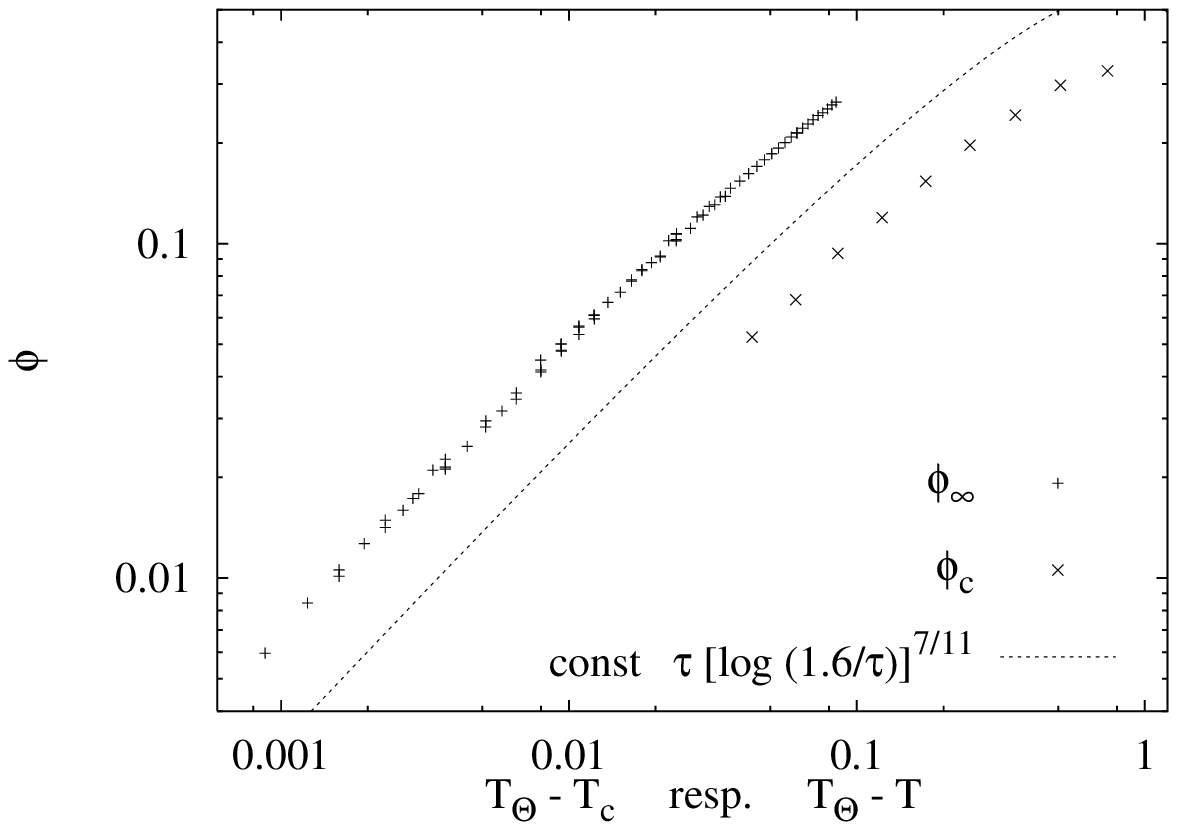,scale=0.85}
\end{center}
\vglue -5mm
\caption{\small Double logarithmic plots of the infinite-$N$ monomer 
density $\phi_\infty$ and of the critical densities for finite $N$, 
as functions of $T_\Theta-T$ resp. $T_\Theta-T_c(N)$. 
The values for $\phi_\infty$ are mostly from \cite{perm}, except 
for the points very close to $T_\Theta$. The slight scatter of these 
points reflects the dependence on system size. The dashed line 
corresponds to eq.(\ref{dupl-phi}), with $\tau=(T_\Theta-T)/T$, and 
with $\log\tau$ arbitrarily replaced by $\log 1.6\tau$. }
\end{figure}

This suggests strongly
that the deviation from the mean field exponent $x_2=1/2$ is 
entirely due to finite-$N$ corrections which vanish for $N\to\infty$. 
This is completely consistent with the very large 
non-asymptotic corrections seen for single chains in the limit
$N\to\infty$ in \cite{hegger,perm}. To stress the similarity between 
the critical monomer density $\phi_c(N)$ as a function of $T_c(N)$, 
and the infinite chain density $\phi_\infty$ at the same value of 
the temperature, we plot them both in fig.9. We see nearly parallel 
curves which suggests that indeed both densities scale with the same 
power of $T_\Theta - T$. For the same values of $T$, Flory-Huggins 
theory predicts $\phi_\infty/\phi_c=3$. For the longest chains our 
data give $3.2\pm 0.2$ for this ratio, with a slight tendency to 
increase with $N$. We see the same small curvature in 
both curves, suggesting that a pure power fit might not be appropriate. 
Such a fit would give an exponent $0.75$ to $0.85$, depending on the 
interval used for the fit. 
The dashed line indicates in contrast the prediction of 
eq.(\ref{dupl-phi}). It does not give a perfect fit, but it definitely 
shows the correct trend. In particular, fitting this curve by a pure 
power law would give an exponent $\approx 0.8$ to $0.9$ (again depending 
on the fit interval), while the 
correct power is 1. It seems thus very likely that all deviations from 
mean field behavior seen in figs. 8 and 9 are due to logarithmic 
corrections.

It seems thus likely that Flory-Huggins theory provides a much 
better description of unmixing of long polymer chains than 
previously thought. To check this directly, we tested the ansatz 
eq.(\ref{mean-free}) directly. According to it, the free energy for 
fixed volume and temperature should consist of a term which 
depends only on the monomer density $\phi$, plus a {\it known} 
entropy contribution which also depends on the chain length $N$. 
To test this, we made simulations at the same $T$ ($=3.5631$) and 
in the same volume ($128^3$ sites) for chain lengths $N=1024, 2048,$
and 4096. Within statistical errors, this $T$ is the critical 
temperature for $N=2048$, thus the chains with $N=1024$ are deeply 
in the single-phase domain, while those for $N=4096$ are deeply in 
the two-phase region. This is illustrated in fig.10a. There, 
constants are added arbitrarily, and fugacities are adjusted 
arbitrarily such that the curves are flattest in a qualitative sense. 

\begin{figure}[hb]
\begin{center}
\vglue -8mm
\begin{minipage}[t]{10.0cm}
\epsfig{file=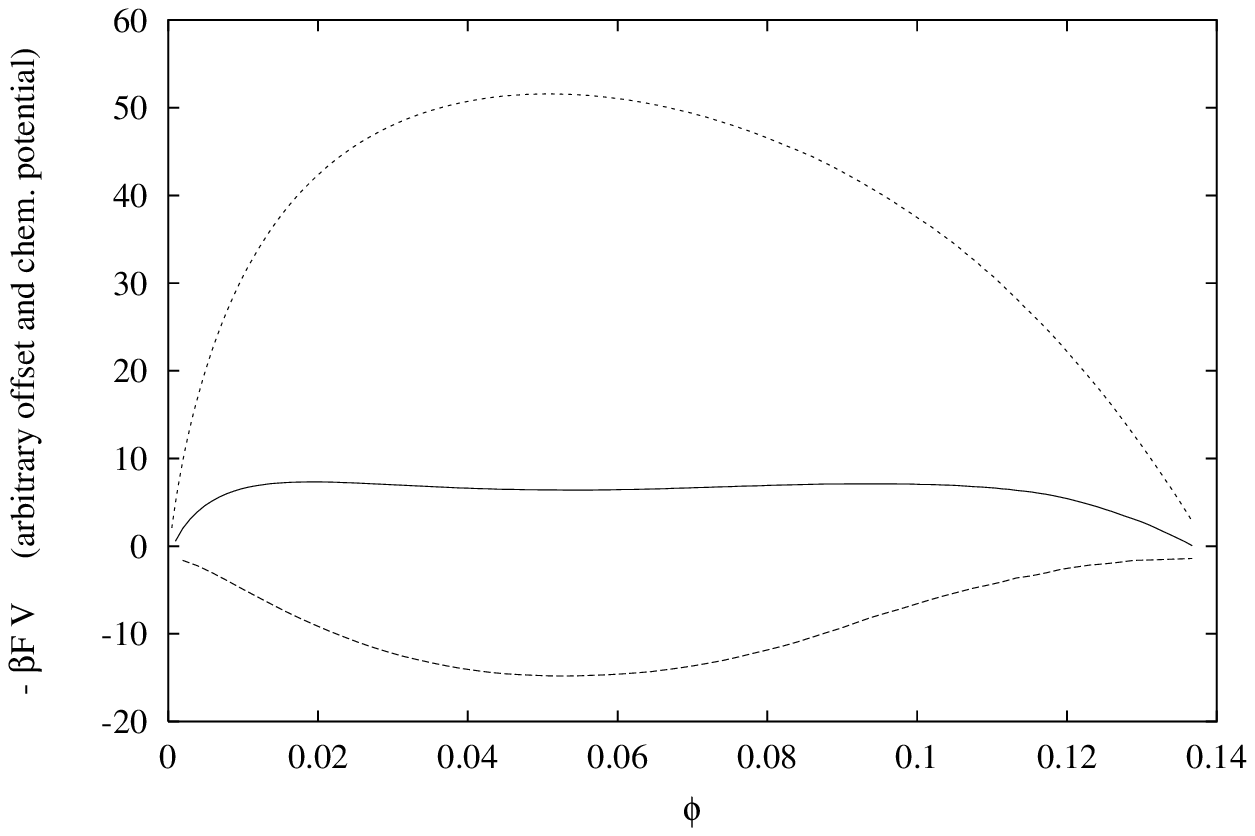,scale=0.80}
\epsfig{file=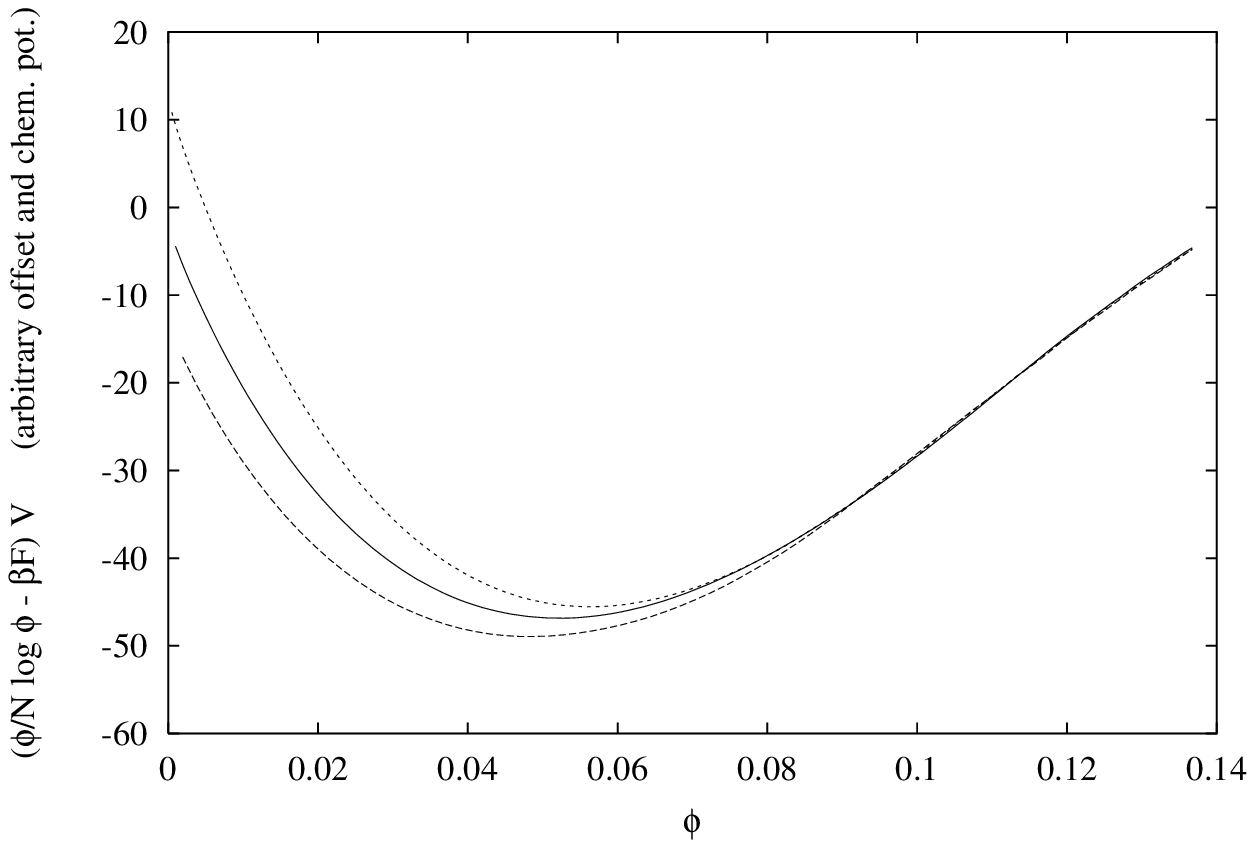,scale=0.80}
\end{minipage}
\end{center}
\caption{\small (a) Total negative free energy $-\beta F V$ of systems 
with three different chain lengths (upper curve: $N=1024$; middle 
curve: $N=2048$; lower curve: $N=4096$) plotted against the monomer 
concentration $\phi$. In all three cases, lattice size and 
temperature were the same: $V=2^{21}$ sites, and $T= 3.5631$. The
latter is close to $T_c$ for $N=2048$, while chains with $N=1024$ 
(4096) are in the single phase (coexistence) domains.
(b) $(\phi/N\log\phi-\beta F)V$ for the same systems as in 
panel a. According to the Flory-Huggins ansatz, this should be 
independent of $N$. Fugacities and additive constants are fixed 
such that the curves coincide for large $\phi$.}
\end{figure}

In fig.10b we show the same data, but after 
removing the supposed entropic contribution $-N^{-1}V\phi\log\phi$. 
In this panel, additive constants and fugacities are adjusted such 
that the curves coincide for large $\phi$. It is there where 
eq.(\ref{mean-free}) should be most reliable: for large monomer 
densities, where chains penetrate substantially, there should be 
hardly any difference between one chain of length $N$ and two chains 
of length $N/2$, up to the entropic difference which is taken out in 
fig.10b. The non-trivial hypothesis underlying eq.(\ref{mean-free}) 
is that the same is true also for small densities. We see from 
fig.10b that it is not perfectly true, but the $N$-dependence in 
fig.10b is much weaker than that in fig.10a. A detailed fit shows 
in addition that none of the curves in fig.10b can be fitted perfectly 
by a cubic polynomial, showing that the internal energy contains 
also terms $\sim \phi^4$ and, since this term has the wrong sign,
higher powers. 

Thus the Flory-Huggins 
ansatz is not exact, but it seems to be a good first approximation.
We should expect deviations from Flory-Huggins due to logarithmic 
corrections, as discussed in sec.1. As pointed out there, the 
leading corrections preserve eq.(2), whence also $\phi_\infty=3\phi_c$ 
should still hold. But
the coefficients of the quadratic and cubic terms, $v$ and $w$, 
should decrease slowly with $N$. This is indeed found when making 
power law fits to the curves in fig.10b (both decrease roughly by 
$20 \%$ when going from $N=1024$ to 4096), but we cannot make a more 
detailed comparison because of the presence of higher than cubic 
terms.

\section{Discussion}

We have applied a novel Monte Carlo scheme to simulate very large 
systems of chain polymers in semidilute solutions. Our system sizes 
are comparable to those of previous analyses as far as chain numbers
are concerned. But our chain lengths are {\it much} longer, extending 
to $>2000$. The latter would have been unfeasible for other 
algorithms we are aware of, and is possible only since our algorithm 
can make use of the fact that long chains close to the critical point 
are nearly free.

Our most solid 
result is that chains at the critical unmixing point are not 
shrunk. They are slightly expanded, but the expansion factor tends 
to a constant for chain length $N\to\infty$. Thus asymptotically, 
for $N\to \infty$, chains are Gaussian in contrast to recent 
speculations, but in agreement with the most recent simulations 
\cite{wmb}. A somewhat less strong result which, however, seems  
also very clear cut, is that the $N$-dependence of the critical 
temperatures is as predicted by Flory-Huggins (mean field) theory. 
Again this is in contrast to recent speculations. 

Strong deviations from mean field behavior were seen in the critical 
density $\phi_c$. Here, a scaling fit would produce the same anomalous 
critical exponent as seen also in experiment. But we show that 
$\phi_c$ can obey this seen anomalous scaling {\it only} if the 
{\it same} anomalous scaling governs also the density inside a 
very large globule, i.e. a single collapsed polymer chain close 
to the $\Theta$ point. For the latter, a superficial analysis also 
suggests anomalous scaling. But theoretical prejudices and more 
careful simulations suggest these might be fake and due to large 
logarithmic corrections to scaling. These corrections are in 
qualitative agreement with field theoretic predictions, and should 
vanish at extremely large chain lengths and extremely close to the 
$\Theta$ point. 

Our simulations do not suggest that deviations from mean field behavior 
--- which must be present because of the anomalous Ising exponents 
--- are ``minimal" in the sense of de Gennes \cite{degennes}: 
in contrast to the prediction $x_1=(1-\beta)/2 = 0.34$ (where $x_1$ 
the exponent for the $N$-dependence of the order parameter) we 
find $x\approx 0.18$, even lower than most phenomenological 
estimates. But we should say that this estimate is by far the most 
shaky of all our results.

It is not clear whether our predictions can be tested by new 
experiments or by re-analyzing old ones. Our chain lengths are 
comparable to those in real experiments ($\approx 1000$ Kuhnian 
lengths), and we predict that mean field behavior should be seen 
in $\phi_c$ only for {\it much} longer chains. This seems not 
feasible at present. On the other hand, fits involving the logarithmic 
corrections might show their presence already for shorter chains. 
Measurements of $T_c(N)$ and of chain dimensions could be improved, 
and should be in agreement with mean field behavior since logarithmic 
corrections seem to be small for them. 

The biggest problem in simulations are the finite sizes of the 
system. In contrast to chain lengths, chain numbers in our 
simulations are orders of magnitude smaller than those 
in real experiments. Nevertheless we believe that finite size 
effects do not seriously affect our conclusions. This is due 
to the use of sophisticated histogram methods \cite{wilding-bruce} 
which allow a detailed comparison with the finite size behavior 
of the Ising model. Essentially they make scaling ansatzes for 
the microcanonical partition sum (a similar method, but without 
the correct finite size dependence of the microcanonical 
partition sum, was proposed in \cite{huller}). We verified this 
dependence partially, which means in particular that we also 
measured the Ising exponents $\beta$ and $\nu$ within our 
simulations. But we found that the simple linear field 
mixing proposed in \cite{wilding-bruce,wmb} does not work 
particularly well. We showed that a nonlinear mixing ansatz 
works much better, but we have no good theoretical reason for 
this ansatz. This is an interesting problem which deserves 
further investigations. But it is not very important as far as 
the precise location of the critical point and the extraction 
of critical parameters are concerned (except for the exponent 
$x_1$), and it cannot affect the above conclusions.

\vspace{.5cm}

{\bf Acknowledgement: }

 We thank many collegues for discussions, mainly 
Gerard Barkema, Ugo Bastolla, Bertrand Duplantier, Walter Nadler, 
and Lothar Sch\"afer. 
We are also indebted to Nigel Wilding for correspondence.

\eject

\end{document}